\documentclass[11pt]{article}
\usepackage[dvips]{graphicx}  
\begin{document}

\title{\LARGE \bf Absolute Motion  and  Quantum Gravity  }  
\author{{Reginald T.
Cahill\footnote{{\bf Process Physics URL:}
scieng.flinders.edu.au/cpes/people/cahill\_r/processphysics.html}}\\
 {School of Chemistry, Physics and Earth Sciences}\\
{ Flinders University }\\ 
{ GPO Box 2100, Adelaide 5001, Australia }\\
{(Reg.Cahill@flinders.edu.au)}}

\date{August  2002}

\maketitle

\begin{abstract}
  A new information-theoretic modelling of reality  has given rise to a quantum-foam description of space,
relative to which  absolute motion is meaningful. In a previous paper (Cahill and Kitto) it was shown that
in this new physics Michelson interferometers show absolute motion effects when operated in dielectric
mode,  as indeed such experiments had indicated, and  analysis of the experimental data showed
that the measured speeds were all consistent with the  Cosmic Microwave Background (CMB) dipole-fit
speed of $369$km/s.  Here   the new physics is applied to the Michelson-Morley 1887 interferometer
rotation curve data to demonstrate that the interferometer data is in excellent agreement with the CMB direction
$(\alpha, \delta)=(11.20^h,-7.22^0)$ as well.  This data also reveals a velocity component caused by the in-flow
of the quantum foam past  the Earth towards the Sun at  $40\pm 15$km/s,  while analysis of the  Miller
interferometer data of 1933 gives $49$km/s, compared to  the theoretical value of $42$km/s. This observed
in-flow is a signature of quantum gravity  effects in the new physics.
\end{abstract}

\vspace{2mm}


\vspace{2mm}

 
  preferred frame, process physics, quantum foam, quantum gravity.

\newpage

\noindent {\bf 1. Introduction} 

 A new information-theoretic modelling of reality known as Process Physics \cite{RC02,RC01} and [14-17]  has
given rise to a quantum-foam description of space, relative to which 
absolute motion is meaningful and measurable. In Ref.\cite{CK} it was shown that
in this new physics Michelson interferometers \cite{Mich} reveal absolute motion  when
operated in dielectric mode,  as indeed   experiments had  indicated, and 
analysis \cite{CK} of the experimental data using the M\'{u}nera \cite{Munera} review of that
 data showed that the measured speeds were  consistent with each other and together
also consistent with the Cosmic Microwave Background (CMB) dipole-fit speed of
$369$km/s \cite{CMB}.   The new physics is here further tested against experiment by
analysing  the Michelson-Morley interferometer rotation data
\cite{MM} of 1887 to demonstrate that  the data is in excellent agreement with the CMB cosmic velocity of the Solar System
through space.  As well as the orbital speed of the Earth the analysis reveals  a quantum-foam in-flow towards the Sun
associated with quantum-gravity effects in the new physics.  So the CMB preferred frame    is detectable in non-microwave
laboratory experiments. These results amount to a dramatic development in fundamental physics.  It is also shown that
analysis of the extensive 1925-1926 dielectric-mode interferometer data  by Miller \cite{Miller2} resulted in an incorrect
direction at $90^0$ to the CMB direction. 

Although  the theory and experiment together indicate that absolute motion is an aspect of
reality one must hasten to note that this theory also implies that the Einstein Special and
General Theory of Relativity formalism remains essentially intact, although the ontology is completely
different.  In
\cite{RC02} it was shown that this formalism arises from the quantum-foam physics, but that the
quantum-foam system  leads to a physically real foliation of the spacetime construct.  Despite this
there are some phenomena which are outside the Einstein  formalism, namely the detection of
absolute motion. We see here the emergence of a new theoretical system which subsumes 
the older theory and covers new phenomena, in particular it unifies  gravity and the
quantum phenomena. 

The new physics provides a different account of the Michelson
interferometer. The main outcome is the presence of the $k^2$ factor in the expression for the time
difference for light travelling via the orthogonal arms 
\begin{equation}\label{eqn:QG1}
\Delta t=k^2\frac{L|{\bf v}_P|^2}{c^3}\cos(2\theta).
\end{equation}
 Here ${\bf v}_P$ is the projection of the absolute velocity ${\bf v}$ of the
interferometer  through the quantum-foam  onto the plane of the interferometer, and $\theta$ is the
angle of one arm relative to ${\bf v}_P$.  The  $k^2$ factor is   $k^2=n(n^2-1)$  where $n$
is the refractive index of the medium through which the light  passes,  $L$ is the
length of each arm and
$c$ is the speed of light relative to the quantum foam. This expression follows from both the
Fitzgerald-Lorentz contraction effect and that the speed of light through the dielectric is
$V=c/n$, ignoring here for simplicity any  drag effects. This is one of the  aspects of the
quantum foam physics that distinguishes it from the Einstein formalism.  The time difference
$\Delta t$ is revealed by the fringe shifts on rotating the interferometer. In Newtonian physics,
that is with no Fitzgerald-Lorentz contraction, $k^2=n^3$, while in Einsteinian physics $k=0$
reflecting the fundamental assumption that absolute motion is not measurable and indeed has no
meaning. So the experimentally determined value of $k$ is a key test of fundamental physics.

Table 1 summarises the differences between the three fundamental theories in their modelling of time, space,
gravity and the quantum, together with their distinctive values for the interferometer parameter $k^2$.  In
particular the Process Physics uses a non-geometric iterative modelling of time in a pre-geometric system
from which a quantum foam description of space is emergent. This quantum foam and quantum matter are
together described by a Quantum Homotopic Field Theory.  Gravity in this modelling is caused by the
inhomgeneous flow of the quantum foam. So Process Physics  is a unification of the quantum and gravity. 
Each theory  subsumes and accounts for the theory above it in the table.  In particular the Einstein
spacetime modelling arises as an approximation to the Process Physics, but with a preferred frame of
reference or foliation. 

\vspace{3mm}
{\small
\hspace{-3mm}\begin{tabular}{| l|c |c|c|c|c|} 
\hline 
{ \bf Theory} &  Time & Space  & Gravity & Quantum &$k^2$ \\
\hline\hline  
{\bf Newton}   & geometry  & geometry & force & Quantum Theory & $n^3$ \\ \hline
 
{\bf Einstein}  & \multicolumn{2}{c}{curved geometry} \vline & curvature & Quantum Field Theory &0
\\ \hline 
{\bf Process}   & process & quantum & inhomogeneous  & Quantum Homotopic  &$n(n^2-1)$
\\
 & &foam & flow & Field Theory & \\ 
\hline
\end{tabular}}

\vspace{2mm}
Table 1: {\small Comparisons of Newtonian, Einsteinian and Process Physics. } 

\vspace{2mm}

 Here we derive  (1) in the new physics and then analyse the Michelson-Morley and Miller
data.  The results reported here are that the small effects (fractional fringe shifts) actually seen by
Michelson and Morley \cite{MM} and by Miller \cite{Miller2}  indicate speeds in agreement with the CMB
speed. This amounts to the observation of absolute motion.  This non-null 
 experimental signature then clearly distinguishes between the three theories in Table 1.

In deriving (1) in the new physics it is essential to note that space is a quantum-foam system
\cite{RC02,RC01}  which exhibits various subtle features. In particular it exhibits real dynamical
effects on clocks and rods. In this physics the speed of light is only $c$  relative to the
quantum-foam, but to observers moving with respect to this quantum-foam the speed appears to be still $c$, but
only because their clocks and rods are affected by the quantum-foam. As shown in \cite{RC02} such observers
will find that records of observations of distant events will be described by the Einstein spacetime
formalism, but only if they restrict measurements to those achieved by using clocks, rods and light
pulses.   It is simplest in the new physics to work in the quantum-foam frame of reference.  If there is a
dielectric present at rest in this frame, such as air, then the speed of light in this frame is
$V=c/n$. If the dielectric is moving with respect to the quantum foam, as in an interferometer attached to the
Earth, then the speed of light relative to the quantum-foam is still $V=c/n$ up to corrections due to 
drag effects.    Hence this new physics requires a different method of analysis from that of the Einstein
physics. With these cautions we now describe the operation of a Michelson interferometer in this new physics,
and show that it makes predictions different to that of the Einstein physics.    Of course experimental
evidence is the final arbiter in this conflict of theories.  

\vspace{3mm}
\noindent  {\bf 2. The Michelson Interferometer}

As shown in Fig.1  the  beamsplitter/mirror when  at $A$ sends a photon $\psi(t)$ into a superposition
$\psi(t)=\psi_1(t)+\psi_2(t)$, with each component travelling in different arms of the interferometer, until
they are recombined in the quantum detector which results in a localisation process, and one spot in the
detector is produced.  Repeating with many photons reveals that the interference between $\psi_1$ and $\psi_2$
at the detector results in fringes.   To simplify the analysis here assume that the two arms are constructed to
have the same lengths $L$  when they are physically parallel to each other and perpendicular to $ v$, so that
the distance  $BB'$ is $L\sin(\theta)$. The Fitzgerald-Lorentz effect in the new physics  is that the
distance  $SB'$  is  $\gamma^{-1} L\cos(\theta)$ where $\gamma=1/\sqrt{1-v^2/c^2}$.  The various other
distances  are $AB=Vt_{AB}$, $BC=Vt_{BC}$, $AS=vt_{AB}$  and $SC=vt_{BC}$, where $t_{AB}$ and $t_{BC}$ are
the travel times.  Applying the Pythagoras theorem to triangle $ABB'$ we obtain
\begin{equation}\label{eqn:1}
t_{AB}=\frac{2v\gamma^{-1}
L\cos(\theta)+\sqrt{4v^2\gamma^{-2}L^2\cos^2(\theta)+4L^2(1-\frac{v^2}{c^2}\cos^2(\theta))(V^2-v^2)}}{2(V^2-v^2)}.
\end{equation}

\vspace{3mm}
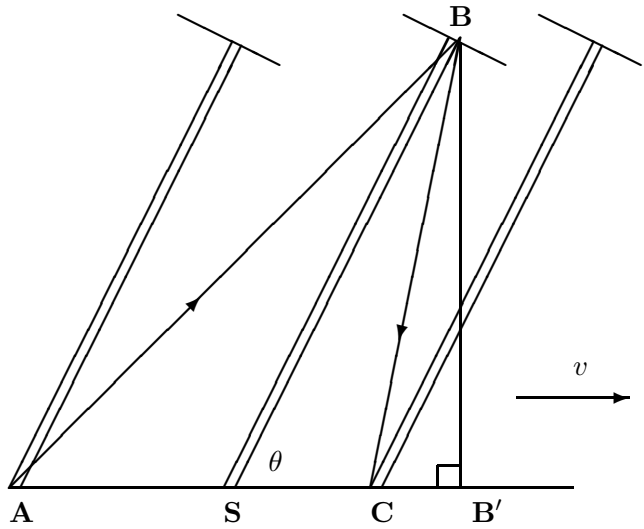
\begin{figure}[t]
\vspace{5mm}\setlength{\unitlength}{1.5mm}
\hspace{45mm}\begin{picture}(20,30)
\thicklines
\put(0,0){\line(1,0){50}}
\put(0,-3){{\bf A}}
\put(32,0){\line(1,5){8}}

\put(32,-3){{\bf C}}
\put(19,0){\line(1,2){20.0}}
\put(20,0){\line(1,2){20}}
\put(19,-3){{\bf S}}
\put(23,+1.5){{$\theta$}}

\put(35,42){\line(2,-1){9}}
\put(39,41){{\bf B}}
\put(0,0){\line(1,1){39.6}}
\put(50,10){{$ v$}}
\put(45,8){\vector(1,0){10}}

\put(0,0){\line(1,2){19.8}}
\put(1,0){\line(1,2){19.6}}
\put(15,42){\line(2,-1){9}}

\put(32,0){\line(1,2){19.8}}
\put(33,0){\line(1,2){19.6}}
\put(47,42){\line(2,-1){9}}

\put(40,0){\line(0,1){39.5}}
\put(41,-3){{\bf B}$'$}

\put(38,0){\line(0,1){2}}
\put(38,2){\line(1,0){2}}

\put(15,15){\vector(1,1){2}}
\put(35.02,15){\vector(-1,-4){0.5}}

\end{picture}

\vspace{3mm}
\caption{\small{One arm  of a Michelson Interferometer travelling  at  angle $\theta$ and   velocity 
${\bf v}$, and shown at three successive times: (i) when photon leaves beamsplitter at $A$, (ii) when photon
is reflected at mirror $B$, and (iii) when photon returns to beamsplitter at $C$. The line $BB'$ defines
right angle triangles $ABB'$ and $SBB'$.  The second arm is not shown but has angle $\theta+90^o$ to 
${\bf v}$. Here ${\bf v}$ is in the plane of the interferometer for simplicity.}}
\end{figure}
\noindent The expression for $t_{BC}$ is the same except for a change of sign of the  $2v\gamma^{-1}
L\cos(\theta)$ term, then  
\begin{equation}\label{eqn:2}
t_{ABC}=t_{AB}+t_{BC}=\frac{\sqrt{4v^2\gamma^{-2}L^2\cos^2(\theta)+4L^2(1-\frac{v^2}{c^2}\cos^2(\theta))
(V^2-v^2)}}{(V^2-v^2)}.
\end{equation}
The corresponding travel time $t'_{ABC}$ for the orthogonal arm  is obtained from  (\ref{eqn:2}) by the
substitution $\cos(\theta) \rightarrow \cos(\theta+90^0)=\sin(\theta)$. The difference in travel times
between the two arms is then $\Delta t= t_{ABC}-t'_{ABC}$. Now trivially $\Delta t =0$  if $v=0$, but 
also $\Delta t =0$ when
$v\neq 0$ but only if $V=c$.  This then would  result in a null result on rotating the apparatus.  Hence the null
result of  Michelson interferometer  experiments in the new physics is only for the special case of
photons travelling in vacuum for which $V=c$.    However if the interferometer is immersed
 in a gas then $V<c$ and a non-null effect is expected on rotating the apparatus, since now 
$\Delta t \neq 0$.  It is essential then in analysing data to correct for this refractive index effect.
For $V=c/n$ we  find for $v << V$  that 
\begin{equation}\label{eqn:4}
\Delta t= Ln(n^2-1)\frac{v^2}{c^3}\cos(2\theta)+\mbox{O}(v^4),
\end{equation}
that is $k^2=n(n^2-1)$,  which gives $k=0$ for vacuum experiments ($n=1$). 

 However if the data from dielectric mode interferometers is (incorrectly)  analysed not using the
Fitzgerald-Lorentz contraction, then, as done in the old analyses,   the estimated Newtonian-physics time
difference is  for $v << V$

\begin{equation}\label{eqn:5}
\Delta t = Ln^3\frac{v^2}{c^3}\cos(2\theta)+\mbox{O}(v^4),
\end{equation}
that is $k^2=n^3$. The value of $\Delta t$ is deduced from analysing the fringe shifts, and then    the
speed
$v_{M}$ (in previous Michelson interferometer type analyses) has been extracted  using (\ref{eqn:5}), instead
of the correct form (\ref{eqn:4}). 
$\Delta t$ is typically of order $10^{-15}s$  in gas-mode interferometers, corresponding to a fractional fringe
shift.   However it is very easy to correct for this oversight.  From (\ref{eqn:4}) and (\ref{eqn:5}) we
obtain, for the corrected absolute speed $v$ through space, and for $n \approx 1^+$, 
\begin{equation}\label{eqn:6}
v=\frac{v_{M}}{\sqrt{n^2-1}}.
\end{equation}
  
Of the early interferometer experiments  Michelson and Morley
\cite{MM} and Miller \cite{Miller2}  operated in air ($n=1.00029$), while that of  Illingworth
\cite{Illingworth} used Helium ($n=1.000035$). We expect then that for air interferometers
$k_{air}^2=0.00058$ (i.e.  $k_{air}=0.0241$) and for Helium $k_{He}^2=0.00007$, which explains why these
experiments reported very small but nevertheless non-null and so significant effects.
 All non-vacuum experiments gave
$k>0$, that is, a non-null effect. All vacuum ($n=1$) interferometer experiments, having $k=0$, give null
effects as expected,  but such experiments cannot distinguish between the new physics and the Einstein physics,
only dielectric-mode interferometers can do that. The notion that the Michelson-Morley experiment gave
a null effect is a common misunderstanding that has dominated  physics for more than a century. By
``null effect'' they meant that the effect was much smaller than expected, and the cause for this is
only now apparent from the above. When the air and Helium interferometer data were re-analysed using
the appropriate $k$ values in \cite{CK} they gave consistent values which were also consistent with
the CMB speed.  So these early interferometer experiments did indeed reveal absolute motion, and
demonstrated that $k\neq 0$. Of the  interferometer experimentalists only Miller consistently argued
that absolute motion had been detected, but failed to convince the physics community.

\vspace{3mm}
\noindent{\bf The Michelson-Morley 1887 Experiment}

 Michelson and Morley reported \cite{MM} that their interferometer experiment in 1887  gave a
``null-result'' which  since then, with rare exceptions, has been claimed to  support the Einstein
assumption that absolute motion has no meaning.  However to the contrary  the Michelson-Morley published
data \cite{MM} shows non-null effects, but much smaller than they expected.  They made observations of
thirty-six  $180^0$ turns  using an $L=11$ meter length air-interferometer in Cleveland
(Latitude $41^0  30'$N) with six turns at 
$12\!:\!\!00$ hrs ($7\!\!:\!\!00$ hrs ST) on each day of July 8, 9 and 11, 1887  and similarly at
$18\!:\!\!00$ hrs ($13\!\!:\!\!00$ hrs ST) on July 8, 9 and 12, 1887. 
The fringe shifts were extremely small but within their observational capabilities.  The  best
$12\!:\!\!00$ and $18\!:\!\!00$ hr rotation data  are shown in Table 2. The dominant effect was a uniform
fringe drift caused by temporal temperature effects on the length of the arms.  After correcting for this the best 
fringe shifts for two   $180^0$ turns are shown in Fig.2. The $18\!:\!\!00$ hr data on July 9 data is
particularly free of observational and vibrational errors, and was used here for detailed fitting.

\vspace{3mm} 
\hspace{20mm}{\small \begin{tabular}{|c|c|c|c|c|c|c|c|c|c|}
\hline
  &16& 1& 2& 3& 4& 5& 6& 7& \\ & 8  &  9  & 10 &11 &12 & 13 & 14 & 15 & 16 \\
\hline
 12:00 &27.3 & 23.5 & 22.0 & 19.3& 19.2& 19.3& 18.7& 18.9 &\\ & 16.2 
&14.3& 13.3& 12.8& 13.3& 12.3& 10.2& 7.3& 6.5  \\ 
\hline
 18:00 &26.0& 26.0& 28.2& 29.2& 31.5& 32.0& 31.3& 31.7& \\ & 33.0
& 35.8& 36.5& 37.3& 38.8&
41.0& 42.7& 43.7& 44.0 \\
\hline
 \end{tabular}}

\vspace{2mm}
\noindent Table 2: {\small  Fringe shift micrometer readings every $22.5^0$of rotation of the  
Michelson-Morley interferometer\cite{MM} for July 11 12:00 hr and July 9 18:00 hr. The arms are at $45^0$ to the 
stone slab supporting base whose orientation is indicated by the marks $16,1,2,..$. North is mark 16. Subtracting
in each case a fit to a+bk,
$\{k=0,1,2,.., 16\}$ removes fringe drifts caused by small uniform temporal temperature changes.  Then multiplying by
$0.02$ for micrometer thread calibration and division by $2$ to get fringe shift per arm  gives the fringe-shift
data points  in Fig.2, but using only the better quality 1st half rotation data.}

\vspace{2mm}
\begin{figure}[ht]
\hspace{25mm}\includegraphics[scale=1.5]{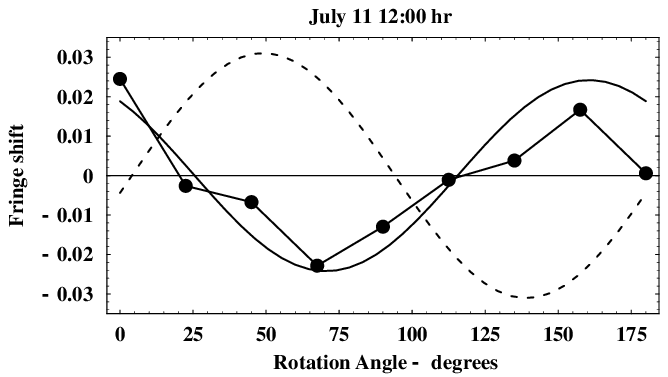}
\vspace{3mm}

 \hspace{25mm}\includegraphics[scale=1.5]{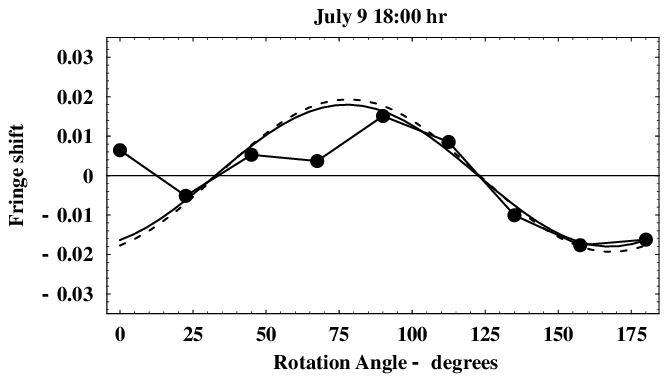}
\vspace{-3mm}
\caption{\small  Data points show the 1887 Michelson-Morley   fringe shifts for $12\!\!:\!\!00$ hrs 
 on July  11  and $18\!\!:\!\!00$ hrs  on July  9  as interferometer was rotated through 180 degrees. The full
curves  come from the quantum-foam theory  best fit to the $18\!\!:\!\!00$ hrs  data. The  theory curves are
$\frac{0.4}{30^2}k^2_{air}v_P^2\cos(2(\theta-\psi-45^0))$, where $v_P$ and $\psi$ are from Table 3 and the
$45^0$ is the offset described in Table 2. The coefficient $0.4/30^2$ arises as the apparatus would give a
$0.4$ fringe shift with $k=1$ if $v_P=30$ km/s \cite{MM}.   The CMB data  gives plots
barely distinguishable from this best fit so long as ${\bf v}_{in}$ and ${\bf v}_{tangent}$ are included. 
The dashed curves shows analogous results using the Miller direction for
${\bf v}_{cosmic}$, which is in  clear disagreement with the
$12\!\!:\!\!00$ hr data. In the best fit to  $18\!\!:\!\!00$ hr data points at $\theta=0^0$ and $67.5^0$  were
neglected.} 
\end{figure} 

In the new physics there are four main velocities that contribute to the total velocity  ${\bf v}$:
\begin{equation}\label{eqn:QG2}
{\bf v}= {\bf v}_{cosmic} +{\bf v }_{tangent} -{\bf v}_{in}-{\bf v}_E.
\end{equation}
Here ${\bf v}_{cosmic}$ is the velocity of the Solar system relative to the cosmological quantum-foam
reference frame, ${\bf v }_{tangent}$ is the tangential orbital velocity of the Earth about the Sun, and 
${\bf v}_{in}$ is  a quantum-gravity radial in-flow  of the quantum foam past the Earth towards the Sun. 
Fig.3a shows  ${\bf v }_{tangent}$ and  ${\bf v}_{in}$. The corresponding quantum-foam in-flow into the
Earth is ${\bf v}_E$ and makes no contribution to a horizontally operated   interferometer.    For circular
orbits the speeds $v_{tangent}$  and  $v_{in}$ are given by \cite{RC02} 
\begin{equation}\label{eqn:QG5}
v_{tangent}=\sqrt{\displaystyle{\frac{GM}{R}}},\end{equation}  
\begin{equation}\label{eqn:QG4}
v_{in}=\sqrt{\displaystyle{\frac{2GM}{R}}},\end{equation}    
where $M$ is the mass of the Sun, $R$ is the distance of the Earth from the
Sun, and $G$ is Newton's gravitational constant. $G$ is essentially a measure of the rate at
which matter effectively `dissipates' the quantum-foam. The gravitational acceleration arises from inhomogeneities in the flow
and is given by ${\bf g}=({\bf v}_{in}.{\bf  \nabla }){\bf v}_{in}$ in this quantum-foam flow physics
\cite{RC02}. These expressions give $v_{tangent}=30$km/s and  $v_{in}=42.4$km/s.

\begin{figure}[t]
\vspace{10mm}
\begin{minipage}[t]{35mm}
\hspace{5mm}\setlength{\unitlength}{1.10mm}
\begin{picture}(0,20)
\thicklines
\put(25,10){\vector(1,0){15}}
\put(25,10){\vector(0,-1){29.5}}
\put(25,10){\vector(1,2){15}}
\qbezier(0,0)(25,20)(50,0)
\put(30,-15){\Large $\bf v_{in}$}
\put(45,10){\Large ${\bf v}_{tangent}$}
\put(40,30){\Large ${\bf v}_{R}$}
\end{picture}

\vspace{22mm}\makebox[20mm][c]{(a)}
\end{minipage}
\begin{minipage}[t]{80mm}
  
\vspace{-45mm}\hspace{40mm}\includegraphics[scale=0.9]{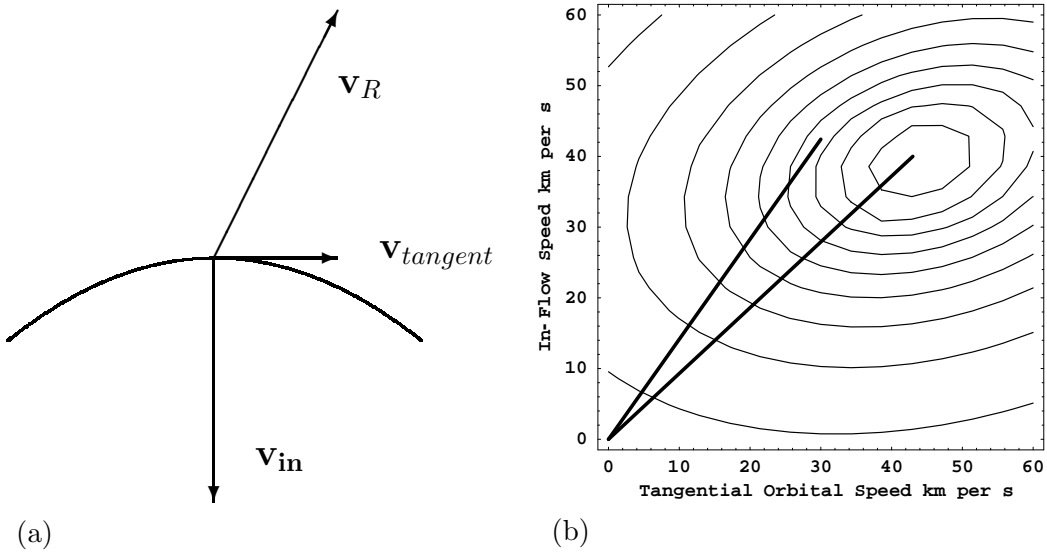}
\makebox[90mm][c]{(b)}
\end{minipage}
\caption{\small  (a) Orbit of Earth defining plane of the ecliptic with tangential orbital velocity ${\bf
v}_{tangent}$ and quantum-foam in-flow velocity  ${\bf v}_{in}$. Then ${\bf v}_{R}={\bf v}_{tangent}-{\bf
v}_{in}$ is the velocity of Earth relative to the quantum foam, after subtracting ${\bf v}_{cosmic}$.  (b)
Corresponding to (a) is determination of best fit to 1887 data for ${\bf v}_{R}$  giving   $|{\bf v}_{in}|=40\pm15$
km/s compared to theoretical value of $42.4$ km/s. Firm lines show ${\bf v}_R$ for best fit and for theory.    } 
\end{figure}

\vspace{10mm}
 {\small
\hspace{0mm}\begin{tabular}{|l|c|c|c|c|r|c|} 
\hline
{ \bf Direction} & $v_{c}$ (km/s)  &Sidereal Time & $v$(km/s)& $v_{p}$  (km/s)& $\psi$(deg.)
&  $v_M$(km/s) \\
\hline 
{\bf CMB:} $(\alpha,\delta)$  & 369.0 &  07:00 July 11   &322.4  & 316.6 & +114.2$^0$ & 7.63
\\
$=(11.20^h,-7.22^0)$   &   & 13:0  July 09 & 323.9  & 269.3  & -151.3$^0$ & 6.49\\  \hline 
{\bf MM1887:} $(\alpha,\delta)$  & 369.0 &  07:00 July 11   & 318.6  & 309.7 & +115.5$^0$ & 7.46
\\
$=(11.20^h,-7.22^0)$   &   & 13:06  July 09 & 324.1  & 271.3  & -149.7$^0$ & 6.53\\  \hline

{\bf Miller:} $(\alpha,\delta)$   &  369.0 &  07:00 July 11  & 366.8 & 348.1  & +4.2$^0$ &
8.39 \\
$=(17.00^h,+70^0)$   &  &   13:00  July 09 & 366.8 & 274.3  & +32.1$^0$  & 6.61 \\
\hline
\end{tabular}}

\vspace{2mm}
\noindent Table 3: {\small Comparisons of  interferometer projected speeds  $v_P$ and azimuths $\psi$ 
corresponding to the total speed $v=|{\bf v}|$,\  where \  ${\bf v}={\bf v}_{c}+{\bf
v}_{tangent}-{\bf v}_{in}$, \ \ for a cosmic speed $v_{c}$  in the  direction indicated by the
celestial coordinates $(\alpha,\delta)$. The azimuth $\psi$ is the angle of ${\bf v}_P$  measured from the  local
meridian ($\pm$ from N). The rows labelled MM1887 refer to the best fit to the nominally $18\!\!:\!\!00$ hr ($13\!\!:\!\!06$ hr,
with a 6 minute offset) Michelson-Morley data from varing $|{\bf v}_{in}|$ and  $|{\bf v}_{tangent}|$, while in rows labelled CMB
the theoretical values for $|{\bf v}_{in}|$ and  $|{\bf v}_{tangent}|$ were used.   $v_M=k_{air}v_p$ is the speed that would be
extracted from the  data using the Newtonian expression (\ref{eqn:5})\footnote{That this  $v_M$ is considerably smaller than the
Earth's orbital speed of  30km/s caused Michelson and Morley to incorrectly report their ``null-result''. This is now understood 
to be a spurious argument.}. The corresponding fringe shifts for MM1887 and Miller as interferometer is rotated are shown in Fig.2.}
 
\vspace{2mm}

Because of limited  data   the direction and magnitude of ${\bf v}_{cosmic}$  was taken as known and  
a least squares fit to the  data by varying $|{\bf v}_{in}|$ and  $|{\bf v}_{tangent}|$ was undertaken.  The results are
shown in Fig.3(b) and  in Table 3, and the fit is graphed in Fig.2.   The  fit is in excellent agreement with the
data and we conclude that ${\bf v}_{cosmic}$ from the interferometer is the same as ${\bf v}_{CMB}$.  Hence the
absolute motion detection capabilities of the Michelson interferometer are clearly evident when used in conjunction
with the new physics.   In finding the best fit we obtain that the magnitude of ${\bf v}_{in}$ is
$40\pm15$ km/s which is consistent with the theoretical value of $42$ km/s.  Fig.3b and Fig.4 clearly
show the determination of  ${\bf v}_{in}$.  This shows that the quantum-foam in-flow effect  is
established and gives us the first signature of quantum gravity effects in the new physics.  Table 3
also shows the various interferometer parameters using the CMB velocity  and theoretical values for $|{\bf v}_{in}|$ and 
$|{\bf v}_{tangent}|$.  

Miller reported
\cite{Miller2} in 1933 a different direction and magnitude for   ${\bf v}_{cosmic}$.  That direction
is at $90^0$ to the CMB/MM direction and is clearly inconsistent with the $12\!:\!\!00$ hr 
Michelson-Morley rotation curve in Fig.2, but it does agree with the $18\!:\!\!00$ hr data.  This
incorrect analysis resulted from the intrinsic $90^0$ directional ambiguity of the interferometer if
continuity of the phase is not carefully followed during a day\footnote{The Miller data was analysed in  \cite{RC03}.
It  now appears that Miller failed to carefully track the diurnal changes in the azimuth $\psi$. Around 11:00 hrs
sidereal time there is a rapid change in $\psi$, and this was not detected by Miller.  
 }.   
\vspace{0mm}
\begin{figure}[t]
\hspace{35mm}\includegraphics[scale=0.9]{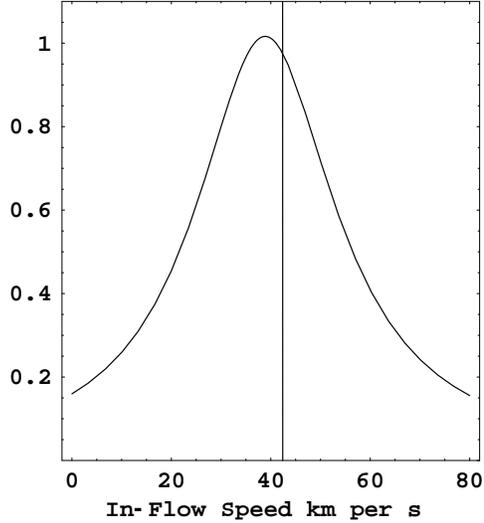}
\vspace{0mm}
\caption{\small{ Plot of reciprocal of relative mean square error for fit to  Michelson-Morley data
versus the quantum-foam in-flow speed giving  $v_{in}= 40\pm 15$km/s compared to theory of 42.4km/s (vertical line).
This is a cut through Fig.3b at fixed $v_{tangent}$, and clearly shows the quantum-gravity in-flow effect.}}
\end{figure}

Nevertheless Miller's extensive Mt.Wilson air-interferometer data  with $L=64$ m 
is capable of confirming some of the above results.  Miller reported in
\cite{Miller2}  particular  observations  over four days in 1925/26 recording the time variation of
the projection ${\bf v}_P$ of the velocity  ${\bf v}$ onto the interferometer throughout each of these
days.     Miller's idea was that ${\bf v}$ should have only two components: (i) a cosmic velocity of the
Solar system through space, and (ii) the orbital velocity of the Earth about the Sun. Over a year this
vector sum  would result in a changing ${\bf v}$, as was in fact observed.  Further,
since the  orbital speed was known, Miller was able to extract from the data the magnitude and
direction of ${\bf v}$ as the orbital speed offered an absolute scale.   Miller was led to the
conclusion that for reasons unknown  the interferometer did not indicate true values of
$v_P$, and for this reason  he introduced the parameter $k$ (we shall denote his values by  $\overline{k}$).
Miller noted, in fact, that $\overline{k}^2<<1$. Fitting  his  data  Miller found $\overline{k}=0.046$ and
$v=210$km/s and the  direction shown in Table 3.  However that $\overline{k} > k_{air}$ now 
confirms that another velocity component has been overlooked.   Miller only knew of the
tangential  orbital speed of the Earth, whereas the new physics predicts that as-well there is a
quantum-gravity radial in-flow ${\bf v}_{in}$ of the quantum foam.
We can re-analyse  Miller's data to extract the speed of this in-flow component.  We easily find that it
is $v_R=\sqrt{v_{in}^2+v^2_{tangent}}$ that sets the scale and not $v_{tangent}$, and so we obtain that the value
of $v_{in}$ implied by  $\overline{k}>k_{air}$  is given by
\begin{equation}\label{eqn:QG3}
v_{in}=v_{tangent}\sqrt{\displaystyle{ \frac{\overline{k}^2}{k_{air}^2}-1 }}
\end{equation} 
 Using the above $\overline{k}$  value and the value of $k_{air}$  we obtain $v_{in}=49$ km/s, which is again in
good agreement with the theoretical value of $42$ km/s.
 Since it is $v_R=\sqrt{3}v_{tangent}$ and not $v_{tangent}$ that sets the
scale we must re-scale Miller's value for $v$ to be $\sqrt{3}\times 210=364$km/s, which now compares favourably
with the  CMB   speed.  Hence  Miller did indeed observe absolute motion as he claimed
but again, as for the Michelson-Morley data,   the quantum gravity in-flow effect is required  in the analysis.

So  the Michelson-Morley  experiment actually amounted to the first quantum gravity
experiment, and the ability of dielectric-mode interferometers to measure absolute motion made this possible.

\vspace{0mm}
\begin{figure}[ht]
\hspace{35mm}\includegraphics[scale=1.0]{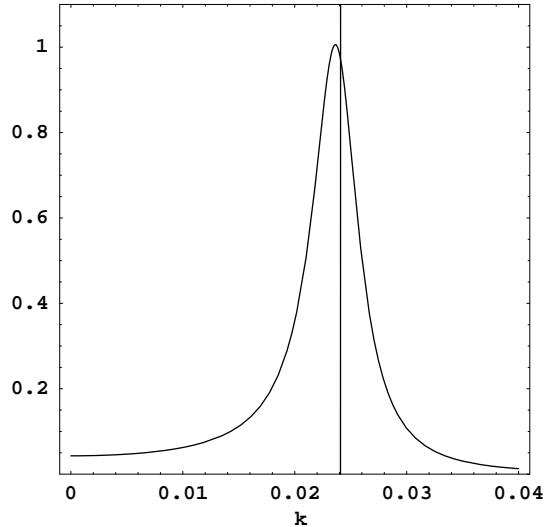}
\vspace{0mm}
\caption{\small{ Plot of reciprocal of relative mean square error for fit to Michelson-Morley data as 
$k$ only is varied with ${\bf v}={\bf v}_{CMB}+ {\bf v}_{tangent} -{\bf v}_{in} $ fixed. 
Best value from comparison is  $k=0.02363$, compared to $k_{air}=0.02410$ (vertical line).}}
\end{figure} 

In Fig.5 is shown best value for $k$ if  ${\bf v}$ is fixed at 
${\bf v}_{CMB}+ {\bf v}_{tangent} -{\bf v}_{in}$ (with ${\bf v}_{tangent}$ and  ${\bf v}_{in}$ set to
theoretical values)  in fit to data, giving  $k=0.02363$ compared to
$k_{air}=0.02410$.  This corresponds to
$n=1.00028$ compared to $n_{air}=1.00029$, demonstrating that the refractive index of air may be
extracted  from the Michelson-Morley data when all three major components of ${\bf v}$ are included.
The results here and above all show that $k \neq 0$.

\vspace{10mm}
\noindent {\bf Conclusions}

The various dielectric-mode interferometer experiments  were never null and their data can now be fully analysed
within the new physics. This analysis  reveals various aspects of the new  quantum-foam phenomena.  The incorrect
reporting by Michelson and Miller of a ``null effect'' was based on using the Newtonian value of $k=1$ and on
$v$ being atleast $30$km/s due to the Earth's orbital motion, and so predicting fringe shifts 10 times larger than
actually seen (the true value for 
$v^2$ in (\ref{eqn:QG1}) is some $10^2$  larger but  the dielectric effect gives a reduction of approximately 
$1/1000$).  Of course that Michelson and Morley saw any effect is solely due to the presence of the air in their
interferometer.   Vacuum interferometer experiments of the same era by Joos \cite{vacuum} gave  
$v_M<1$km/s, and are consistent with a null effect as  predicted by the quantum-foam physics. 
If Michelson and Morley had more carefully reported their results the history of
physics over the last 100 years would have been totally different. 

The experimental results analysed herein and in \cite{CK} show that absolute motion is detectable.  
This is motion with respect to  a quantum-foam system that is space. As well quantum matter
effectively acts as a sink for the quantum-foam, and the flow of that quantum-foam towards the Sun has
been confirmed by the data.  These results are in conflict with the fundamental assumption by Einstein that
absolute motion has no meaning and so cannot be measured.  Vacuum interferometer experiments do give null results,
for example see  \cite{vacuum,  KT, BH, Muller}, but they only check the Lorentz contraction effect, and this is
common to both theories. So they are unable to distinguish the new physics from the Einstein physics.  As well
that the interferometer experiments and their results fall into two classes, namely vacuum and dielectric has
gone unnoticed. The non-null results from dielectric-mode interferometers have always been rejected on the 
grounds that they would be in conflict with the many successes of the Special and General Theory of
Relativity.   However this is not strictly so, and it turns out that these successes survive in the new
physics, which actually subsumes the Einstein formalism, even though the absolute motion effect is not in the
Einstein physics. Einstein essentially arrived at a  valid formalism from a wrong assumption.  The new more
encompassing process physics [1-3, 14-17] allows the determination of a physically real foliation of the spacetime
construct (the Panlev\'{e}-Gullstrand foliation)  and so it actually breaks the diffeomorphism symmetry of General
Relativity.

 The author thanks Warren Lawrance for on-going discussions of new-generation dielectric-mode interferometer
experiments.

\end{document}